\documentstyle[tighten,preprint,aps,prb]{revtex}  
\begin{document}
\draft
\title{Anisotropy and temperature dependence of the optical conductivity 
in ${\rm \bf LaMnO_3}$}
\author{K. H. Ahn \cite{Ahn} and A. J. Millis } 
\address{Department of Physics and Astronomy, Rutgers University
\\
Piscataway, New Jersey 08854}
\maketitle
\begin{abstract}
A tight binding parameterization of the band 
structure, along with a mean field treatment of Hund,
electron-electron,
and electron-lattice couplings, 
is used to obtain the 
full optical conductivity tensor of 
${\rm LaMnO_3}$
as a function of temperature.
We predict striking changes with temperature 
in the functional form and magnitude of
the optical absorption.
Comparison of our results to data will 
determine the Hund, electron-lattice, and 
electron-electron interactions.
\end{abstract}
\narrowtext

\newpage
\section{Introduction}
In recent years the colossal magnetoresistance 
manganese perovskites have attracted much attention.\cite{Royal}
The low energy ($\hbar\omega < $4 eV) physics of these materials is 
believed to be governed by Mn $e_g$ electrons,
which 
are coupled by a strong Hund's coupling, $J_{\rm H}$, to
Mn $t_{2g}$ symmetry `core spins'
and also interact with each other and with lattice distortions.
Three important and still unresolved issues are the value of $J_{\rm H}$, 
the strength of the Jahn-Teller(JT) electron-phonon coupling 
$\lambda$, 
and the magnitude of the effective $e_g$-$e_g$ on-site Coulomb 
interaction $U$ governing the low energy physics. 
The relative importance of the electron-electron and 
electron-phonon interactions is 
a particularly important  
open question.

Hund's coupling $J_{\rm H}$ is believed to be large,
because in good samples at low temperatures, observed magnetic moments 
are close to their
saturation values.\cite{Wollan55}
However, $J_{\rm H}$ has not been directly measured.
Estimates 
ranging from $2 J_{\rm H}S_{\rm c} \approx$ 2 eV
(Ref. \onlinecite{Okimoto97})
to 3 eV 
(Ref. \onlinecite{Quijada98})
have been obtained from the analysis of the optical measurements
of doped materials.
However, these estimates depend upon 
proper identification
of features in the observed optical conductivity,
and this identification 
is at present disputed.

The value of the effective Coulomb repulsion is also unclear.
High energy photoemission spectroscopy\cite{Saitoh95}
yields a bare repulsion $U_{\rm bare}\approx$ 8 eV,
but also implies that the manganites are in the class of
charge-transfer materials,\cite{Zaanen85} 
as are the high-$T_c$ cuprates.
In charge-transfer materials 
the low energy ($\hbar\omega \ll U_{\rm bare}$)
electronic degrees of freedom are complicated linear combinations of 
transition metal and oxygen orbitals, 
and the effective repulsion for these orbitals 
may be much less than $U_{\rm bare}$.
For example, in high $T_c$ materials, $U_{\rm bare} \approx$ 10 eV  
but the $U$ relevant to the low energy physics is 
rather less;
perhaps $U\approx$ 2 eV (Ref. 
\onlinecite{Hybertsen89}).
The proper value of the $U$ relevant 
to the low energy ($\hbar\omega <$ 4 eV) physics of 
manganites
is at present disputed.
Some authors argue that electron-electron interactions
are dominant; others that electron-phonon interactions are crucial.

In this communication, we show how measurements of the magnitude, 
anisotropy and 
temperature dependence of the optical conductivity 
of the manganites' insulating `parent compound' ${\rm LaMnO_3}$
can help to resolve these issues.
We obtain an effective Hamiltonian for the low energy degrees
of freedom 
by fitting band theory calculations
to a tight binding model
and adding Coulomb interaction terms.
We then
determine its optical conductivity
as a function of temperature, 
and show
how the different features in the spectrum may be used to determine the 
interaction parameters.
The advantage of the tight binding approach
is that it compactly represents the low energy degrees of freedom,
allows straitforward inclusion
of interaction effects,
and is computationally cheap enough that many parameter
choices may be investigated.

The material is a (0,0,$\pi $) antiferromagnet 
at low temperatures, 
while  
a paramagnet at room temperature.
The Hund's coupling therefore leads to a peak structure with a pronounced 
and strongly temperature-dependent anisotropy.
Observation of this effect would provide a definitive determination of the 
Hund's coupling parameter.

The Jahn-Teller and Coulomb interactions lead to 
similar effects in the absorption
spectrum;
indeed the two interactions roughly add.
The strength of the Jahn-Teller coupling may be determined 
from crystallography data or from band theory;
interestingly the two determinations agree very well.
The effect of the Jahn-Teller coupling on the optical conductivity 
may essentially be calculated  exactly.
We compare the results of this calculation to the optical 
conductivity data,
and argue that differences are due to the Coulomb interaction.
To estimate the strength of the Coulomb interaction 
we use a Hartree-Fock approximation
and calculate the optical conductivity of the Jahn-Teller plus Coulomb 
system.
The optical conductivity of ${\rm LaMnO_3}$
was previously calculated by Solovyev {\it et al.}\cite{Solovyev96}
and by Terakura {\it et al.}\cite{Terakura99}
using the LDA method,
but they did not consider the temperature dependence,
or present information about
the dependence
of the observed conductivity
on parameters.
When we calculate using parameters tuned to give similar
band gaps,
we find the maximum conductivity 
is factor of 2-3 larger than the result in Ref.
\onlinecite{Solovyev96}
and 4 times than in Ref. \onlinecite{Terakura99}.
The difference in absorption
strength is not understood at present,
but is discussed in more
detail in Section V,
which shows that the error is unlikely to
arise from the tight binding approach.

The rest of this paper is organized as follows:
Section II presents the Hamiltonian, band structure, 
and the parameter determination,
Section III gives the formulation
of the optical conductivity,  
calculation results, and the comparison with experiments,
Section IV describes the effect of the Coulomb interaction,
Section V 
discusses the accuracy of the approximation we have used,
and Section VI
is the conclusion.

\section{Hamiltonian and band structure}  
At and below room temperature,
${\rm LaMnO_3}$ exists in a distorted form of the $AB{\rm O_3}$ 
perovskite structure.
According to band theory 
calculations\cite{Satpathy96.PRL,Satpathy96.JAP,Pickett96}
the 
conduction band is derived mainly from Mn $e_g$ symmetry d-orbitals 
and is well
separated from other bands.
We find that the band structure appropriate 
to the ideal cubic $AB{\rm O_3}$ 
perovskite structure may be well represented 
by the following tight binding model.
\begin{equation}
H_{\rm KE}+H_{\mu}=-\frac{1}{2}\sum_{\vec{i},\vec{\delta},a,b,\alpha} 
t^{ab}_{\vec{\delta}} 
 d^{\dagger}_{\vec{i}a\alpha} d_{\vec{i}+\vec{\delta}b\alpha}
+H.c.
-\mu\sum_{\vec{i},a,\alpha} d^{\dagger}_{\vec{i}a\alpha}  
d_{\vec{i}a\alpha} .  
\end{equation}
Here
$\vec{i}$ represents the coordinates of the Mn sites
(which in the ideal structure are arranged in a simple cubic lattice),
$a$,$b$ represent the two degenerate Mn $e_g$ orbitals
on a site,
$\delta(=\pm x,y,z)$
labels the nearest neighbors of a Mn site,
$\alpha$ denotes the spin state, 
and $t^{ab}_{\vec{\delta}}$
is the hopping amplitude between orbital $a$ 
on site $\vec{i}$ and $b$ on site $\vec{i}+\vec{\delta}$.
We choose  $|\psi_1>=|3z^2-r^2>$ and  $|\psi_2>=|x^2-y^2>$
as the two linearly independent $e_g$ orbitals on a site. 
The hopping matrix $t^{ab}_{\vec{\delta}}$ has a special form:
for hopping along the $z$ direction, it connects only the two $|3z^2-r^2>$ 
states, thus
\begin{eqnarray}
t_z=t_{-z}=t_o \left(  \begin{array}{cc}
                                   1   &   0  \\
                                   0   &   0 
                        \end{array}     \right).
\end{eqnarray}
The hopping matrices in the other bond directions are obtained 
by appropriate
rotations and are
\begin{eqnarray}
t_x&=&t_{-x}=t_o \left(  \begin{array}{cc}
                                   1/4   &  - \sqrt{3}/4 \\
                       - \sqrt{3}/4   &   3/4 
                        \end{array}     \right),  \\
t_y&=&t_{-y}=t_o \left(  \begin{array}{cc}
                                   1/4   &   \sqrt{3}/4 \\
                        \sqrt{3}/4   &   3/4 
                        \end{array}     \right) . 
\end{eqnarray}  
As noted in the introduction, there is
a substantial high energy ($U_{\rm bare} \approx$ 8 eV) 
photoemission evidence 
for strong on-site Coulomb
interactions in the manganites, 
which  places them in the class of `charge transfer'
materials.
The relevance of the band theory calculation may therefore be questioned.
We argue, however, that the effects of the interactions 
at the low($\hbar\omega <$ 4 eV)
energies of interest may be determined by comparing the predictions 
of the band theory 
calculation to data;
the results we present will allow this comparison to be made.
The high $T_{\rm c}$ superconductors provide an instructive example.
These are also charge transfer insulators 
with a very large high energy on-site repulsion
$U_{\rm high}$ (Ref.
 \onlinecite{Hybertsen89,Anderson97} ).  
The low energy excitations are complicated objects 
called Zhang-Rice singlets,
but it has been established that the effective interaction relevant to the 
low energy theory is much less than 
$U_{\rm high}$, and that band theory(albeit with a renormalized hopping)
describes the electron dispersion well.
We therefore suggest that band theory is an appropriate starting point 
in the manganite case as well.

We now turn to the electron-lattice coupling.
Below 800 K, ${\rm LaMnO_3}$ exists in a distorted form 
of the $AB{\rm O_3}$ perovskite structure.
The important distortion is a Jahn-Teller distortion 
in which Mn-O bond lengths
are changed in such a way that if we choose a Mn site as an origin,
the distortion is even-parity and the sum of the lengths 
of the 6 Mn-O bonds
is unchanged.
This distortion lifts the degeneracy of the $e_g$ levels on a site.
To represent this we define  
$u_{\vec{i}}^a$ as the  
displacement along the {\it a} direction
of an oxygen ion located between
Mn ions at $\vec{i}$ and $\vec{i}+\hat{a}$,
and 
we define
$v^a_{\vec{i}}=u^a_{\vec{i}}-u^a_{\vec{i}-\hat{a}}$.
The Jahn-Teller distortion term 
may then be written as
\begin{equation}
H_{\rm JT}=-\lambda\sum_{\vec{i}\alpha} \left( \begin{array}{c}
           d^{\dagger}_{1,\vec{i},\alpha}   \\
           d^{\dagger}_{2,\vec{i},\alpha} 
            \end{array}    \right)^T
         \left(\begin{array}{cc}
          v^z_{\vec{i}}-\frac{1}{2}(v^x_{\vec{i}}+v^y_{\vec{i}})   &    
         \frac{\sqrt{3}}{2} (v^x_{\vec{i}}-v^y_{\vec{i}})    \\
          \frac{\sqrt{3}}{2} (v^x_{\vec{i}}-v^y_{\vec{i}})   &
        -v^z_{\vec{i}}+\frac{1}{2}(v^x_{\vec{i}}+v^y_{\vec{i}}) 
            \end{array}     \right)
           \left( \begin{array}{c}
            d_{1,\vec{i},\alpha}   \\
             d_{2,\vec{i},\alpha} 
            \end{array}    \right).
\end{equation}
The experimentally observed distortion has two components:
a $Q_2$ type staggered distortion
with wave vector $(\pi,\pi,0)$, and
a $Q_3$ type uniform distortion.
This distortion leads to a Jahn-Teller term of the form
\begin{equation}
H_{\rm JT}=-\lambda\sum_{\vec{i},\alpha} \left( \begin{array}{c}
           d^{\dagger}_{1,\vec{i},\alpha}   \\
          d^{\dagger}_{2,\vec{i},\alpha} 
             \end{array}    \right)^T
         \left(\begin{array}{cc}
       -\bar{v}   &    
      (-1)^{i_x+i_y} \bar{w}    \\
      (-1)^{i_x+i_y} \bar{w}  &
           \bar{v} 
        \end{array}     \right)
        \left( \begin{array}{c}
       d_{1,\vec{i},\alpha}   \\
       d_{2,\vec{i},\alpha} 
      \end{array}    \right),
\end{equation}
where $\bar{w}$
and 
$\bar{v}$
are the amplitudes of the staggered($Q_2$) 
and uniform($Q_3$) distortion respectively.

We next consider the Hund's coupling.
This leads to a term 
\begin{equation}
H_{\rm Hund}=\sum_{\vec{i},a,\alpha} J_{\rm H}\vec{S_{\rm c}^i} \cdot 
d^{\dagger}_{\vec{i},a,\alpha}
\vec{\sigma}_{\alpha\beta}
d_{\vec{i},a,\beta},
\end{equation}
where $\vec{S_{\rm c}^i}$
represents the $t_{2g}$ core spin and
$\vec{\sigma}$
the Pauli matrix. 
At $T$ = 0 K, the magnetic structure is 
of a (0,0,$\pi$) 
antiferromagnet; leading to
\begin{equation}
H_{\rm Hund}=J_{\rm H}S_{\rm c}
\sum_{\vec{i},a} \left[ \left(1-(-1)^{i_z} \right) 
 d^{\dagger}_{\vec{i},a,\uparrow}  d_{\vec{i},a,\uparrow} +
\left(1+(-1)^{i_z} \right) 
 d^{\dagger}_{\vec{i},a,\downarrow}  d_{\vec{i},a,\downarrow} \right].
\end{equation}
The total Hamiltonian is the sum of the terms considered so far.
\begin{equation}
H_{\rm tot}=H_{\rm KE}+H_{\mu}+H_{\rm JT}+H_{\rm Hund} \label{eq:tot}.
\end{equation}
Fourier transforming into $k$ space 
results in the  expression for $H_{\rm tot}$
given in the Appendix A.
By diagonalizing this matrix, we can find the energy levels.

At $T$ = 0 K, the unit cell is doubled twice, once by spin
and once by orbital ordering.
We have two orbital states 
for each of the two spin states, and four Mn sites
per unit cell. 
Due to the symmetry between two spin states,
we will have two-fold degeneracy
for each level.
Therefore, we have eight separate bands.
The ground state is obtained by
filling the energy levels below the chemical potential $\mu$, 
which is determined to give the correct number of electrons 
per unit cell.
We denote the energy levels by 
$E_j(\vec{k})$ ($j$=1,2,...,8) in the order of the increasing energy.
The band structure determined by Eq. (\ref{eq:tot})
is shown in Fig. 1 for the parameters which 
provide the best fit 
to the published band calculations.\cite{Satpathy96.JAP}
Crudely speaking,
the bands fall into 4 pairs, which
may be understood by setting $t_o=0$
(as occurs at $(\pi/2,\pi/2,\pi/2)$);
in this case 
we have four separate energy levels on each site:
which are $E_{1,2}=-\lambda \sqrt{\bar{v}^2+\bar{w}^2}$, 
$E_{3,4}=\lambda \sqrt{\bar{v}^2+\bar{w}^2}$,
$E_{5,6}=2 J_{\rm H}S_{\rm c} - \lambda \sqrt{\bar{v}^2+\bar{w}^2}$, and 
$E_{7,8}=2 J_{\rm H}S_{\rm c} + \lambda \sqrt{\bar{v}^2+\bar{w}^2}$.
When we have non-zero $t_o$, these levels split and become dispersive.
In the low temperature (0,0,$\pi$) antiferromagnetic structure, the 
Hund's coupling suppresses the $z$-directional hopping;
the bands thus become more two-dimensional 
as  $J_{\rm H}S_{\rm c}$ increases. 

Band theory calculation is used to determine 
the three parameter values of our model Hamiltonian: 
$t_0$, $\lambda$, and $J_{\rm H}S_{\rm c}$.
To find these, we fit our band structure calculation
to the LDA band calculation for the JT distorted 
${\rm LaMnO_3}$ by Satpathy {\it et al.} \cite{Satpathy96.JAP}
at high symmetry points in reciprocal space,
($\pi$,0,0), (0,0,0), ($\pi/2$, $\pi/2$,  $\pi/2$), and ($\pi$,0,$\pi/2$).
The standard deviation is $\approx$ 0.2 eV, 
and maximal error of 0.4 eV
occurs at  ($\pi/2$, $\pi/2$,  $\pi/2$) for 
the lower JT level of the upper Hund state, $E_{5,6}$.
The determined parameter values are 
$t_o$=0.622 eV, $\lambda$=1.38 eV/$\AA$, 
and $2 J_{\rm H}S_{\rm c}$=2.47 eV.
The fitted band structure is shown in Fig. 1.
These parameters fit the LDA band calculations 
for the JT distorted and buckled actual ${\rm LaMnO_3}$ structure published
by Satpathy {\it et al.} \cite{Satpathy96.PRL}
with a similar size of error.
The above $t_o$ and $J_{\rm H}S_{\rm c}$ 
are similar to the values obtained by Myrasov {\it et al.}\cite{Myrasov98}
from an LDA calculation for the ideal cubic structure.

This parameter $\lambda$ may be independently determined 
by fitting the observed lattice distortion \cite{Ellemans71} 
to a simple model 
of localized electrons which are Jahn-Teller-coupled to
a harmonic lattice as 
explained in Ref.
\onlinecite{Millis96}. 
This reference shows that the amplitude 
of the observed distortion fixes the parameter 
$\lambda/(K_1 a_0)$,
where $K_1$ is the spring constant between the nearest 
neighbor Mn-O pairs
and $a_0$ is the average Mn-O distance.
The following two equations are derived in Ref.
 \onlinecite{Millis96}.
\begin{eqnarray}
a_o e^{a}&=&\frac{\lambda}{K_1+2 K_2}
\left[ \cos 2\left( \theta_1+\psi^a \right) + 
\cos 2 \left( \theta_2 +\psi^a \right) \right]  \label{eq:uniform} \\
u_s^a &=& 
\frac{\lambda}{2K_1} \left[ \cos 2 \left( \theta_1 +\psi^a \right) 
- \cos 2 \left( \theta_2 + \psi^a \right) \right] \end{eqnarray}
where 
$e^a$($a=x,y,z$) is the uniform strain, $u_{\rm s}^a$ is the staggered 
oxygen displacement,
$K_2$ is the nearest neighbor Mn-Mn spring constant,
$\psi^{x,y,z}=-\pi/3,\pi/3,0$, and $\theta_1$, $\theta_2$ 
parameterize the orbital states on the two sublattices:
$|\psi>=\cos \theta |3z^2-r^2>+ \sin \theta |x^2-y^2> $.
(Eq. (\ref{eq:uniform}) corrects a factor of two error in Eq. (10) of Ref.
\onlinecite{Millis96}.)
Using the relation $\theta_2=\pi-\theta_1$ derived in Ref.
\onlinecite{Millis96},
we get the following expression of $\lambda/(K_1 a_0)$.
\begin{equation}
\frac{\lambda}{K_1 a_0}=\sqrt{\left(e^{z}\frac{1+2K_2/K_1}{2}\right)^2+
\left(\frac{2 u_s^x}{\sqrt{3} a_0}\right)^2}
\end{equation}
From Ellemans {\it et al.}'s results,\cite{Ellemans71} $e^{z}$=-0.0288, 
$u_s^x$=0.141 $\AA$,
and
$a_0$=4.034 $\AA$.
For $0 \leq K_2/K_1 \leq 1.0$, we obtain 
$0.0428 \leq  \lambda/(K_1 a_0) \leq 0.0591$.
$K_1$ is estimated from the frequency 
of the highest lying bond stretching mode
measured in this material by Jung {\it et al.}. \cite{Jung98}
The measured bond stretching mode has a peak at 70.3 meV.
From the relation 
$(\hbar \omega)^2=2K_1(m_{\rm Mn}^{-1}+m_{\rm O}^{-1})$,
we obtain
$K_1$ =7.36 eV/$\AA^2$.
Using $a_0$ = 4.034 $\AA$,
the range of $\lambda$ obtained is between 1.27 eV/$\AA$ and 1.76 eV/$\AA$,
which includes the value obtained above.
We can, in fact, determine the lower bound of $K_2/K_1$ 
from the structural transition temperature as explained 
in Ref.
 \onlinecite{Millis96}.
In Ref.
\onlinecite{Millis96}, 
the mean field estimation of the structural phase transition 
temperature was found to be 
$T_{\rm s}^{\rm MF}=3\lambda^2 K_2/[ 2 K_1 (K_1+2 K_2)]$ = 750 K = 65 meV.
Considering that mean field theory overestimates 
the transition temperature,
we obtain 
\begin{equation}
\frac{3}{2}\frac{\lambda^2 K_2}{K_1 (K_1+2 K_2)} > 65 \ {\rm meV}.
\end{equation}
Combining the previous expression of $\lambda/(K_1 a_0)$ versus $K_2/K_1$,
we can determine the range of $K_2/K_1$.
The determined range is $K_2/K_1 > $ 0.26,
and gives $\lambda >$ 1.36 eV /$\AA$,
which is remarkably close to the value obtained by band fitting.

\section{Optical conductivity}
\subsection{At T= 0 K}
Optical conductivity per volume, $\sigma$, can be found in the following
way. 
The electromagnetic field couples to the electrons 
via the Peierls phase factor 
$t_{\delta} \rightarrow t_{\delta} 
\exp \left( 
 i e \vec{A} \cdot \vec{\delta} a_0 
 / \hbar
\right)$.
This implies the total current operator 
\begin{equation}
\hat{\vec{J}} = \frac{\delta H}{\delta \vec{A}}
 = -\frac{i e a_0}{2 \hbar}
\sum_{\vec{i},\vec{\delta},a,b,\alpha} 
t^{ab}_{\vec{\delta}} \vec{\delta} \left[ 
\exp \left( \frac{ i e \vec{A} \cdot \vec{\delta} a_0 }
  { \hbar }  \right)
 d^{\dagger}_{\vec{i}a\alpha} d_{\vec{i}+\vec{\delta}b\alpha}
- H.c. \right].
\end{equation}
By expanding about $\vec{A}$, we obtain 
\begin{equation}
\hat{\vec{J}} = \hat{\vec{J_p}} + \hat{\vec{J_d}} + O(A^2),
\end{equation}
where
\begin{eqnarray}
\hat{\vec{J_p}} &=& 
 -\frac{i e a_0}{2 \hbar}
\sum_{\vec{i},\vec{\delta},a,b,\alpha} 
t^{ab}_{\vec{\delta}} \vec{\delta} \left( 
 d^{\dagger}_{\vec{i}a\alpha} d_{\vec{i}+\vec{\delta}b\alpha}
- H.c. \right),  \\
\hat{\vec{J_d}} &=&
 \frac{e a_0}{2 \hbar}
\sum_{\vec{i},\vec{\delta},a,b,\alpha} 
t^{ab}_{\vec{\delta}} \vec{\delta} 
\frac{ e \vec{A} \cdot \vec{\delta} a_0 }
  { \hbar }
\left( 
 d^{\dagger}_{\vec{i}a\alpha} d_{\vec{i}+\vec{\delta}b\alpha}
+ H.c. \right). 
\end{eqnarray}
Since 
$
\vec{J}(\vec{A}) = 
<0(\vec{A}) | \hat{\vec{J_p}} + \hat{\vec{J_d}} | 0(\vec{A}) >
= -i \omega \Sigma \vec{A} + O(A^2)
$,
where 
$\Sigma = \int dV \sigma $ 
is the total conductivity,
by expanding $\vec{J}(\vec{A})$
up to linear order in $\vec{A}$, we can find $\sigma$.
Since $\hat{\vec{J_d}}$ is already linear in $\vec{A}$,
it gives the following contribution in $\sigma $.
\begin{equation}
\sigma_d^{\lambda \nu}
=-\frac{1}{ i \omega} \left( \frac{e a_0}{\hbar} \right)^2
K_{\lambda \lambda} \delta_{\lambda \nu} \frac{1}{a_0^3}
\end{equation}
where
\begin{equation}
K_{\lambda \lambda}=\frac{1}{N_{Mn}}
<0| \frac{1}{2} \sum_ 
{\vec{i},\vec{\delta}=\pm \hat{\lambda}, a, b, \alpha } 
t^{ab}_{\vec{\delta}}  \left( 
 d^{\dagger}_{\vec{i}a\alpha} d_{\vec{i}+\vec{\delta}b\alpha}
+ H.c. \right) 
|0>. 
\end{equation}
Since $\hat{\vec{J_p}}$ does not contain $\vec{A}$, we should 
use higher order perturbation to obtain linear term.
Second order perturbation yields
\begin{equation}
\sigma_p^{\lambda \nu} = - \frac{1}{i \omega N_{Mn} a_0^3}
\sum_{n} 
\frac{ <0|J_{p \lambda}^{\dagger} | n> < n | J_{p \nu} | 0 > }
{\hbar \omega - (E_n - E_0 ) + i \epsilon },
\end{equation}
where 
$\epsilon$ is an infinitesimal 
introduced to make the expression well defined. 
The above procedure is a standard linear response theory,
which is also used  
in Ref. \onlinecite{Millis90},
and explained in detail in Ref. \onlinecite{Dagotto94}.

By transforming into energy eigenbasis, we can 
obtain explicit expression of $K_{\lambda \lambda}$
and $\sigma_p $, which gives $\sigma$. 
The results are
\begin{equation}
K^{\lambda \lambda} = \frac{2}{(2 \pi)^3}
\int_{R} d \vec{k}
\sum_{E_j(\vec{k}) < \mu} 
2\cos k_a B_a(\vec{k})_{jj}  \label{eq:KE}
\end{equation}
and 
\begin{equation}
(\sigma_p)_{aa}= - \frac{2}{i \omega a_o^3} 
\frac{1} {(2 \pi)^3} \int_{R} d\vec{k}
\sum_{ E_j(\vec{k})<\mu , E_{j'}(\vec{k})>\mu}
\frac{\left|
(e a_0 / \hbar ) 2 \sin k_a B_a(\vec{k})_{jj'}
\right|^2 }
     { \hbar \omega-E_{j'}(\vec{k})+E_j(\vec{k}) + i \epsilon }, 
\end{equation} 
where 
\begin{eqnarray}
B_a(\vec{k})&=&M(\vec{k})^{\dagger} T_a M(\vec{k}), \\
T_x&=&\left( \begin{array}{cccc}
                     t_x &       0  &   0    & 0 \\
                       0   &  -t_x &    0  &  0  \\
                       0   &  0   & t_x & 0  \\
                     0  &  0  &  0  &  -t_x   \end{array} \right), \\
T_y&=&\left( \begin{array}{cccc}
                     t_y & 0 & 0 & 0  \\
                      0  &  -t_y  &  0  &  0  \\
                     0  &  0 & t_y &  0  \\
                      0  &  0  &  0  &  -t_y \end{array} \right), \\
T_z&=&\left( \begin{array}{cccc}
                     t_z & 0 & 0 & 0  \\
                      0  &  t_z  &  0  &  0  \\
                     0  &  0 & -t_z &  0  \\
                      0  &  0  &  0  &  -t_z \end{array} \right),
\end{eqnarray}
$M(\vec{k})$ is the matrix diagonalizing $H_{\vec{k},\uparrow}$ 
in the Appendix A,
$a_o$ is the distance between Mn ions, 
and
$R$ is defined in the Appendix A.
The factor of 2 comes from the two spin states.
Therefore, the real part of $\sigma$ is 
\begin{equation}
Re[\sigma_{aa}]=\frac{2}{\omega a_o^3} 
\frac{1}{(2 \pi)^3} \int_{R} d\vec{k}
\sum_{ E_j(\vec{k})<\mu , E_{j'}(\vec{k})>\mu}
\left|
\frac{e a_0}{\hbar} 2 \sin k_a B_a(\vec{k})_{jj'}
\right|^2
\frac{\epsilon}
     {[\hbar \omega-E_{j'}(\vec{k})+E_j(\vec{k})]^2+\epsilon^2}. 
\end{equation} 

The total oscillator strength in the couductivity of our Hamiltonian
may be obtained as follows. \cite{Millis90,Kohn64}
Optical conductivity $\sigma $ can be written
as
\begin{equation}
\sigma^{\lambda \nu} (\omega ) =
\left( \frac{e^2}{\hbar^2 a_0} 
K_{\lambda \nu} \delta_{\lambda \nu} - \chi_{\lambda \nu}
\right)
/ ( -  i \omega ),
\end{equation}
where 
\begin{equation}
\chi_{\lambda \nu} = 
\frac{1}{N_{Mn} a_0^3}
\sum_{n} 
\frac{<0|J_{p\lambda}^{\dagger}|n><n|J_{p\nu}|0>}
{\hbar \omega - (E_n - E_0) + i \epsilon }.
\end{equation}
At large $\omega$, $\chi_{\lambda \nu} \propto 1/ \omega $. 
Therefore, $\lim_{\omega \rightarrow \infty} 
\sigma^{\lambda \nu} (\omega )= 
\frac{e^2}{\hbar^2 a_0} K_{\lambda \lambda} \delta_{\lambda \nu}
/(- i \omega)$.
Large $\omega$ limit of the Kramers-Kronig relation 
\begin{equation}
Im[ \sigma (\omega )] = -( 2 \omega / \pi )
P \int_0^{\infty} ds \frac{  Re[ \sigma (s)]}{ s^2 - \omega^2 }
\end{equation}
yields the following sum rule:
\begin{equation}
K^{\lambda \lambda}=
\frac{\hbar^2 a_0}{e^2}
\frac{2}{\pi}
\int_0^{\infty} d \omega Re[ \sigma_{\lambda \lambda} (\omega )]. 
\label{eq:sum} 
\end{equation}

From crystallography studies 
in Ref.  
\onlinecite{Ellemans71},
 $\bar{w}$=0.488 $\AA$ and $\bar{v}$=0.174 $\AA$ 
at $T$ = 0 K.
Figures 2(a), 2(c), and 2(e) show the $T$=0 K 
optical conductivities $\sigma_{xx}$ and $\sigma_{zz}$
calculated for three values of the coupling constant $\lambda$
with $t_o$ and $J_{\rm H}S_{\rm c}$
predicted by band theory and 
$\bar{v}$ and $\bar{w}$ from crystallographic data.
Figure 2(a) shows $\sigma_{xx}$ and $\sigma_{zz}$ for the case 
$2 \lambda \sqrt{\bar{v}^2+\bar{w}^2} < 2 J_{\rm H}S_{\rm c}$.
For $\sigma_{xx}$ (solid line), we see a large peak 
at the Jahn-Teller splitting,
corresponding to motion within one plane.
Note the jump in absorption at the gap edge,
characteristic two-dimensional feature,
a weak feature at $2 J_{\rm H}S_{\rm c}$, corresponding to electron
trajectories which overlap from one plane to the next, 
and an extremely weak feature at
the sum of the Jahn-Teller and Hund's splitting.
For $\sigma_{zz}$(dotted line), we see a very weak feature at
the Jahn-Teller energy, corresponding to 
a small amplitude for an electron to tunnel through an intervening
plane and land on a `correctly oriented' core spin,
a large peak at the Hund's energy, and
another peak at the sum of 
Hund's and Jahn-Teller energies. 
The sharp peak at the Hund's energy in $\sigma_{zz}$
originates from the essentially parallel
bands seen in Fig. 1 between $( \pi /2, \pi /2, \pi /2 )$
and $(\pi, 0, \pi /2 )$. 
In LSDA band calculation in Ref. \onlinecite{Satpathy96.JAP},
these two bands are not exactly parallel, but deviates by 0.34 eV.
Therefore, we expect this will cause the broadening
of the peak by $\epsilon$=0.17 eV.
The effect of this broadening is further discussed in Section V.
Figure 2(c) shows the case of comparable Hund's and Jahn-Teller couplings.
We see that the structure becomes more complicated as the features
overlap, and the band structure becomes less two-dimensional.
The low energy shoulder starting from 1 eV
originates from the transition
between the opposite spin directions.
Finally, Fig. 2(e) shows the case of the Jahn-Teller 
coupling  greater than Hund's coupling.
In this case the Hund's feature appears strongly
for both $\sigma_{xx}$ and $\sigma_{zz}$, 
whereas the Jahn-Teller feature is 
now almost completely absent in $\sigma_{zz}$.
Figures 3(a), 3(c), and 3(e) show the results for a lower value of 
$J_{\rm H}S_{\rm c}$,
which show similar features as Figs. 2(a), 2(c), and 2(e).

We have also studied the change of the optical conductivity 
caused
by the 
change of the ordered orbital state.
The orbital ordering angle,
$\theta$,
defined in Section II,
is related to the 
lattice distortion 
$\bar{v}$ and $\bar{w}$ by 
\begin{equation}
\cos 2 \theta = - \frac{\bar{v}}{\sqrt{\bar{v}^2+\bar{w}^2}}
\end{equation}
For observed $\bar{v}$ and $\bar{w}$ at $T=0$, $\theta \approx 54^o $.
We have varied the ratio between $\bar{v}$ and $\bar{w}$ 
without changing JT splitting $2\lambda \sqrt{\bar{v}^2+\bar{w}^2}$.
The results are shown in Fig. 4.
When $\theta=\pi/6$, the orbital ordering is $x^2-z^2/y^2-z^2$ type,
and when $\theta=\pi/3$,  $3x^2-r^2/3y^2-r^2$ type.
When $\theta=\pi/4$, orbital state is between
the two configurations, and $\bar{v}=0$, i.e., 
there is
no uniform 
distortion.
As orbital ordering changes from 
$x^2-z^2/y^2-z^2$ to
$3x^2-r^2/3y^2-r^2$,
total spectral weight of $\sigma_{zz}$ has substantially decreased.
If $\theta$ is further varied toward $\theta$=$\pi/2$,
which corresponds to non-staggered 
$x^2-y^2$ type orbital ordering, then the $z$-direction
conduction becomes smaller.
For $\sigma_{xx}$, the spectral weight has moved close to the lower edge 
without appreciable change in the total spectral weight.

\subsection{At T=300 K} 
The model described above is appropriate to describe 
the optical conductivity
when the spin and lattice degrees of freedom are 
perfectly ordered
at $T$=0 K.
If the temperature is non-zero, then the core spins will 
fluctuate from the ground 
state configuration.
In the actual material, 
long range order is lost
at $T_{\rm N} \approx$ 140 K. 
Therefore, by room temperature 
it is reasonable to assume
that the core spins are completely 
disordered.
In principle, similar considerations apply 
to the lattice degrees of freedom,
but because room temperature is much lower than 
the structural transition 
temperature 800 K,
we may neglect lattice fluctuation and assume 
a static JT distortion.

To describe the system at $T_{\rm N}\ll T \ll $ 800 K, 
we develop the effective 
Hamiltonian in the following way.
Instead of choosing spin basis along a fixed direction independent
of sites, 
we choose $\Uparrow$  on site $\vec{i}$ as the direction of $e_g$ electron
parallel to the core spin on site $\vec{i}$, 
and $\Downarrow$ as its opposite direction.
Therefore, the Hund coupling energy is 
\begin{equation}
H_{\rm Hund}=2 J_{\rm H}S_{\rm c}
\sum_{\vec{i},a}  
 d^{\dagger}_{\vec{i},a,\Downarrow}  d_{\vec{i},a,\Downarrow}.  
\end{equation}
$H_{\mu}$ and $H_{\rm JT}$ do not change their forms by the change
of spin basis.
We define the angle between the core spin directions
on site $\vec{i}$ and on site $\vec{i}+\vec{\delta}$ 
as $\theta_{\vec{i},\vec{i}+\vec{\delta}}$,
so that the kinetic energy is given by 
\begin{eqnarray}
H_{\rm KE}&=&-\frac{1}{2}\sum_{\vec{i},\vec{\delta},a,b} 
t^{ab}_{\vec{\delta}} 
\left[ 
\cos \left( \frac{ \theta_{\vec{i},\vec{i}+\vec{\delta}}}{2} \right)
d^{\dagger}_{\vec{i} a \Uparrow} d_{\vec{i}+\vec{\delta}b \Uparrow}+
\cos \left( \frac{ \theta_{\vec{i},\vec{i}+\vec{\delta}}}{2} \right)
d^{\dagger}_{\vec{i} a \Downarrow} d_{\vec{i}+\vec{\delta}b \Downarrow} 
\right.  \nonumber \\
&+&\left. 
\sin \left( \frac{ \theta_{\vec{i},\vec{i}+\vec{\delta}}}{2} \right)
d^{\dagger}_{\vec{i} a \Uparrow} d_{\vec{i}+\vec{\delta}b \Downarrow}+
\sin \left( \frac{ \theta_{\vec{i},\vec{i}+\vec{\delta}}}{2} \right)
d^{\dagger}_{\vec{i} a \Downarrow} d_{\vec{i}+\vec{\delta}b \Uparrow}+
H.c. \right].  
\end{eqnarray}
This new representation of the Hamiltonian reduces to
the previous $H_{\rm tot}$ at $T$=0 K.
At $T \gg T_{\rm N}$, $\theta_{\vec{i},\vec{i}+\vec{\delta}}$ will 
be completely 
random. 
Therefore, 
we argue it is appropriate to average the Hamiltonian by setting 
$< \cos (  \theta_{\vec{i},\vec{i}+\vec{\delta}}/2 ) >=$
$< \sin (  \theta_{\vec{i},\vec{i}+\vec{\delta}}/2 ) >=2/3$,
which gives the following effective kinetic energy term
\begin{eqnarray}
H^{\rm eff}_{\rm KE}&=&-\frac{1}{3}\sum_{\vec{i},\vec{\delta},a,b} 
t^{ab}_{\vec{\delta}} 
\left[ 
d^{\dagger}_{\vec{i} a \Uparrow} d_{\vec{i}+\vec{\delta}b \Uparrow}+
d^{\dagger}_{\vec{i} a \Downarrow} d_{\vec{i}+\vec{\delta}b \Downarrow} 
\right.  \nonumber \\
&+&\left. 
d^{\dagger}_{\vec{i} a \Uparrow} d_{\vec{i}+\vec{\delta}b \Downarrow}+
d^{\dagger}_{\vec{i} a \Downarrow} d_{\vec{i}+\vec{\delta}b \Uparrow}+
H.c. \right].  
\end{eqnarray}
Transforming into $\vec{k}$ space yields the effective total Hamiltonian 
shown in the Appendix B.

At room temperature, the core spin directions are 
fluctuating in time and space,
which broadens the levels.
To incorporate this physics,
we introduce a phenomenological broadening $\Gamma$, 
leading to the following expression of the optical conductivity.
\begin{eqnarray}
& &Re[\sigma_{aa}] \nonumber \\
&=&
\frac{1}{\omega a_0^3}
\frac{1}{(2\pi)^3}
\int_{S} d \vec{k}
\int \int dE dE' 
\sum_{ E_j(\vec{k})<\mu , E_{j'}(\vec{k})>\mu}
\frac{ \epsilon
|
( e a_0 / \hbar ) 2 \sin k_a B'_a(\vec{k})_{jj'}
|^2}
{[\hbar \omega-E'+E]^2+\epsilon^2}
\frac{\Gamma / \pi}{[E'-E_{j'}(\vec{k})]^2+\Gamma^2} \nonumber \\
& &\frac{\Gamma / \pi}{[E-E_j(\vec{k})]^2+\Gamma^2}   \nonumber \\
&=&
\frac{1}{ \omega a_0^3 }
\frac{1}{(2 \pi)^3} \int_S d \vec{k}
\sum_{ E_j(\vec{k})<\mu , E_{j'}(\vec{k})>\mu}
\frac{2 \Gamma 
|
( e a_0 / \hbar ) 2 \sin k_a B'_a(\vec{k})_{jj'}
|^2
}
{[\hbar \omega-E_{j'}(\vec{k})+E_j(\vec{k})]^2+(2 \Gamma)^2}, 
\end{eqnarray}
where
\begin{eqnarray} 
B'_a(\vec{k})&=&M'(\vec{k})^{\dagger} T'_a M'(\vec{k}), \\
T'_x&=&\frac{2}{3}\left( \begin{array}{cccc}
                     t_x &       0  &   t_x    & 0 \\
                       0   &  -t_x &    0  &  -t_x  \\
                       t_x   &  0   & t_x & 0  \\
                     0  &  -t_x  &  0  &  -t_x   \end{array} \right),  \\
T'_y&=&\frac{2}{3}\left( \begin{array}{cccc}
                     t_y & 0 & t_y & 0  \\
                      0  &  -t_y  &  0  &  -t_y  \\
                     t_y  &  0 & t_y &  0  \\
                      0  &  -t_y  &  0  &  -t_y \end{array} \right), \\
T'_z&=&\frac{2}{3}\left( \begin{array}{cccc}
                     t_z & 0 & t_z & 0  \\
                      0  &  t_z  &  0  &  t_z  \\
                     t_z  &  0 &  t_z &  0  \\
                      0  &  t_z  &  0  &  t_z \end{array} \right),
\end{eqnarray}
$M'(\vec{k})$ is the matrix diagonalizing $H_{\vec{k}}^{\rm eff}$ 
in the Appendix B,
$S$ is defined in the Appendix B, and
$t_{x,y,z}$ is the 2$\times$2 matrix defined previously.
$\Gamma$ may be estimated from 
the root mean square fluctuation in the hopping;
we find 
\begin{equation}
\Gamma \approx t_o \sqrt{<\cos^2 (\theta /2)>-<\cos (\theta /2)>^2}
\approx \frac{t_o}{3\sqrt{2}}.
\end{equation}

The general features of 
$\sigma_{xx}$ and $\sigma_{zz}$ at $T  \gg T_{\rm N}$ are these:
Because we have random spin directions along both $x$ and $z$ directions,
both $\sigma_{xx}$ and $\sigma_{zz}$ show JT, Hund, and JT+Hund peaks.
Due to the anisotropy of the lattice distortion, we still expect 
anisotropy in the peak intensity.
The broadening due to random spin directions means the peaks 
become smoother
than  
the $T$=0 K case. 

Optical conductivities calculated for room temperature are shown 
in Figs. 2(b), 2(d), and 2(f).
For this calculation, we use the same $\lambda$, $t_o$, 
and $J_{\rm H}S_{\rm c}$ as 
in Figs. 2(a), 2(c), and 2(e), 
but we use the  room temperature lattice parameters,
which differ slightly from 
the 0 K lattice parameters.
We obtain  $\bar{w}$=0.417 $\AA$ and $\bar{v}$=0.155 $\AA$ from Ref.
\onlinecite{Ellemans71}.
As expected, the peaks are substantially broadened and indeed 
in Figs. 2(d) and 2(f)
only two peaks are visible.
Figures 3(b), 3(d), and 3(f) show similar results, obtained 
for the parameters used in Figs. 3(a), 3(c), and 3(e). 
The upturn of the optical conductivity at around zero frequency is 
an artifact of
the crude consideration of the fluctuation in our model.

We have calculated the variation of spectral weight with
temperature and parameter values;
the results 
obtained by 
Eqs. 
(\ref{eq:KE}) 
and 
(\ref{eq:sum}) 
are consistent, and 
shown in the Table I.
It shows that at $T$=0 K, 
$K_{xx}$ sensitively decreases as $\lambda$ increases, 
but insensitive to $J_{\rm H}S_{\rm c}$,
and $K_{zz}$  decreases  as $\lambda$ or 
$J_{\rm H}S_{\rm c}$ increases,
which can be understood from the spin and lattice configuration at $T$=0 K.
On the other hand, at $T$=300 K, both $K_{xx}$ and 
$K_{zz}$ have moderate dependence 
on both $\lambda$ and $J_{\rm H}S_{\rm c}$, which originates 
from the paramagnetic spin configuration. 
Temperature dependence of the total spectral weight is relatively weak,
even though the spectral weight of each peak depends 
sensitively on temperature.
Particularly, when the Jahn-Teller splitting is the much 
less than the Hund's splitting
(as in Figs. 2(a) and 2(b)), it is possible to identify 
the lowest energy feature at both 0 K and 300 K 
as arising from transitions
between the parallel-spin but different Jahn-Teller states, 
and  to determine the spectral weight in this feature.
When we define 
\begin{equation}
K_{a, JT}=\frac{2 \hbar^2 a_0 }{\pi e^2}
\int_{JT peak} d \omega \sigma_{aa} (\omega ),
\end{equation}
we find 
$K_{x,JT}^{0K}$=0.271 eV,  
$K_{x,JT}^{300K}$=0.151  eV 
for Figs. 2(a) and 2(b),
whose ratio is between 1/2 and 2/3,
as predicted in Ref.
\onlinecite{Quijada98}.
The extra $T$=0 K  JT spectral weight is pulled down from higher 
energy peaks as the spin disorder is decreased.
It is also noteworthy that the peak shape is
more asymmetric at $T$ = 0 K than at $T$ = 300 K,
due to the 2-d character.
 
We now compare our results to data.
This comparison is preliminary because
the available data disagree. 
Optical conductivity for polycrystalline ${\rm LaMnO_3}$ was measured 
at room temperature
by Jung {\it et al.}.\cite{Jung97}
Because the crystal directions are random in polycrystalline samples,
the observed quantity is $\sigma_{\rm av}=2 \sigma_{xx}/3+\sigma_{zz}/3$,
which we have also plotted in Fig. 5
(We assume here that the crystallite size is large.).
Figure 5(a) shows the results for the parameter values determined 
from band fitting.
The data in Ref.
\onlinecite{Jung97}
exhibit two main structures;
a lower peak centered at 1.9 eV with maximum intensity  
420 $\Omega^{-1}{\rm cm}^{-1}$
and $K_{av, {\rm JT}}=2 K_{x, {\rm JT}}/3+
K_{z, {\rm JT}}/3\approx$ 0.115 eV,
and a 
peak centered at around 4.5 eV with a much 
larger intensity.
They attribute the 4.5 eV feature
to $e_g$ - $O_{2p}$ transitions
beyond the scope of our model,
and assign the peak at 1.9 eV to JT-split $e_g$-$e_g$ transitions
within the parallel spin manifold,
which we agree with.
In this interpretation, the transitions to the reversed spin states 
are obscured by Mn-O
transition. 
Recently, room temperature optical reflectivity spectra 
using a cleaved single crystal surface of
${\rm La_{1-x}Sr_xMnO_3}$ have been measured by 
Takenaka {\it et al.}.\cite{Takenaka98}
Although it is referred to as a single crystal, we believe that 
the sample of ${\rm LaMnO_3}$ is 
micro-twined.
In their results, the Jahn-Teller peak appears at around 2.5 eV, 
having maximum intensity around 600 $\Omega^{-1}{\rm cm}^{-1}$
with width around 1.5 eV, corresponding to 
$K_{\rm av, JT} \approx$ 0.141 eV,
and the Mn-O peak appears at 5 eV with maximum intensity 2800 
$\Omega^{-1} {\rm cm}^{-1}$.
Similar results were obtained by Okimoto {\it et al.}.\cite{Okimoto97}
Takenaka {\it et al.}'s results show a weak shoulder at 1.9 eV,
which our model cannot explain.

From Fig. 5(a), one sees that, 
if the Jahn-Teller interaction were the only important one,
the observed lattice distortions would lead to a peak in 
$\sigma_{\rm av}$ at 1.2 eV with maximum intensity 
$\approx$ 1200  $\Omega^{-1}{\rm cm}^{-1}$,
width $\approx $ 1.0 eV,
and $K_{\rm av,JT}=$ 0.223 eV.
The maximum intensity or spectral weight is 
much larger than observed in either experiment and the peak position is
lower.
From Fig. 5(c), we see that 
the data of Ref.
\onlinecite{Takenaka98}
may be approximately modeled by  
use of a stronger electron lattice coupling
(or larger amplitude lattice distortion),
which moves the peak to higher energy
and reduces its spectral weight. 
The data in Ref. \onlinecite{Jung97} are very difficult 
to reconcile with theory,
because as can be seen from Fig. 3(b),
one cannot simultaneously fit the peak amplitude and energy:
choosing interaction parameters to fit the peak position 
leads to an amplitude,
which is too large.
The combination of peak energy and amplitude could only be explained 
if the actual hopping were rather smaller
than the band theory value
(say $t_o \approx $ 0.4 eV rather than 0.6 eV).

Further optical data would be very desirable
(especially measurements at lower $T$).
For the present we assume the Takenaka {\it et al.}'s data are correct,
and consider their interpretation in more detail.
We believe that the combination of band calculation
and estimates from the crystallography data adequately 
fix the magnitude of the Jahn-Teller splitting.
We therefore believe that the differences between 
the Takenaka {\it et al.} data 
and Fig. 5(a) are mainly due to the Coulomb interaction
whose effects we study in the next section.

\section{Coulomb interaction}
We now add an on-site Hubbard-type  
Coulomb repulsion to our Hamiltonian.
Because we have 2 orbital and 2 spin states on each site,
we have in principle 6 different Coulomb repulsion terms,
which may be generally written 
($\hat{n}$ is the density operator)
\begin{equation}
H_{\rm Coulomb}=\sum_{\vec{i}} \sum_{a \neq b} \sum_{\alpha \neq \beta}
U_{(\alpha,\beta),(a,b)} \hat{n}_{\vec{i},\alpha,a} 
\hat{n}_{\vec{i}\beta,b}
\end{equation}

For simplicity we study this Hamiltonian in the Hartree-Fock approximation,
which we believe is reasonably accurate for the simple quantities
(peak position and spectral weight) important for our analysis.
Corrections to the Hartree-Fock approximation are due to 
quantum fluctuations.
We have compared Hartree-Fock to exact results for 
the case of strongest fluctuations,
namely the one-dimensional Hubbard model,
and find agreement to within 30\%
for spectral weight\cite{Stafford92}
(peak positions are not available).
We believe that in the case of present interest,
the combination of three dimensionality,
the large core spins, and the localization due to the electron-phonon
interaction
renders the Hartree-Fock approximation sufficiently accurate.

In the approximation, 
one of the two density operators is replaced by its expectation value,
which is determined self-consistently.
The approximation explicitly breaks symmetry in spin and orbital space,
so the issue of basis choice arises.
We choose the orbital basis picked out 
by the observed lattice distortion and 
the spin basis picked out by the magnetic ordering.
We refer the higher and lower lying orbital states 
by + and $-$ respectively,
and the spin states by $\Uparrow$ and $\Downarrow$ defined previously.
The mean-field Hamiltonian may then be written as  
\begin{equation}
H^{\rm MF}_{\rm Coulomb} = \sum_{\vec{i}} 
U_{\Uparrow +} d_{\vec{i} \Uparrow +}^{\dagger}  d_{\vec{i} \Uparrow +}
+ U_{\Uparrow -} d_{\vec{i} \Uparrow -}^{\dagger}  d_{\vec{i} \Uparrow -}  
+ U_{\Downarrow +} d_{\vec{i} \Downarrow +}^{\dagger}  
d_{\vec{i} \Downarrow +}
+ U_{\Downarrow -} d_{\vec{i} \Downarrow -}^{\dagger}  
d_{\vec{i} \Downarrow -},  \label{eq:HU}
\end{equation} 
where 
$U_{\Uparrow,+}$ = $U_{\Uparrow\Uparrow + -} <\hat{n}_{\Uparrow -}>$+
$U_{\Uparrow\Downarrow + +} <\hat{n}_{\Downarrow +}>$ +
$U_{\Uparrow\Downarrow + -} <\hat{n}_{\Downarrow -}>$, etc..
 
Eq. (\ref{eq:HU})  may be reorganized into a term 
proportional to the total $e_g$ density operator,
which renormalizes the chemical potential and is of no interest,
a term which couples to the total $e_g$ spin operator 
and changes the Hund's coupling, 
and terms which renormalize
the local Jahn-Teller splitting in a manner which differs
for electrons
locally parallel and antiparallel 
to the core spin.  
Therefore, $H_{\rm JT}+H_{\rm Hund}+H^{\rm MF}_{\rm Coulomb}$ 
can be cast into the following form.
\begin{eqnarray}
H_{\rm JT}+H_{\rm Hund}+H^{\rm MF}_{\rm Coulomb}
&=&\sum_{i} \lambda_{\Uparrow}' \sqrt{\bar{v}^2+\bar{w}^2} 
( \hat{n}_{\vec{i} \Uparrow +} -  \hat{n}_{\vec{i} \Uparrow -} )
+ \lambda_{\Downarrow}'  \sqrt{\bar{v}^2+\bar{w}^2} 
( \hat{n}_{\vec{i} \Downarrow +}  
- \hat{n}_{\vec{i} \Downarrow -} ) \nonumber \\
& &+2 J_{\rm H}' S_{\rm c} ( \hat{n}_{\vec{i} \Downarrow +}  
+ \hat{n}_{\vec{i} \Downarrow -} ),
\end{eqnarray}
where 
\begin{eqnarray}
\lambda_{\Uparrow}' &=& \lambda+\frac{U_{\Uparrow +}
-U_{\Uparrow -}}{2 \sqrt{\bar{v}^2+\bar{w}^2}},  \\
\lambda_{\Downarrow}' &=&\lambda+\frac{U_{\Downarrow +}
-U_{\Downarrow -}}{2 \sqrt{\bar{v}^2+\bar{w}^2}},  \\
2 J_{\rm H}' S_{\rm c} &=& 2 J_{\rm H}S_{\rm c} 
+ \frac{1}{2}( U_{\Downarrow +} + U_{\Downarrow -} 
- U_{\Uparrow +} - U_{\Uparrow -} ).
\end{eqnarray}
Because the hopping matrices are defined in $|\psi_1>$ and $|\psi_2>$ 
orbital basis, we need to transform the basis.
For this purpose, it turns out useful to define
\begin{eqnarray}
\lambda_{\Uparrow}'&=& \lambda_{av}' + \delta \lambda' , \\
\lambda_{\Downarrow}'&=& \lambda_{av}' - \delta \lambda' .
\end{eqnarray}
If we transform into $k$ space in $|\psi_1>$ and $|\psi_2>$ basis, 
we obtain the Hamiltonian to calculate optical conductivity.
Details are shown in the Appendix C along with the 
expression of the number operators.

As the simplest case, we consider the case where 
$U_{(\alpha,\beta),(a,b)}=U$, independent of 
$(\alpha,\beta)$ and $(a,b)$.
For this case, we use $t_o$, $\lambda$, 
and $J_{\rm H}S_{\rm c}$ obtained previously 
and determine the values of $U$ by calculating optical conductivity 
at $T$=300 K and comparing with experimental JT peak position.
For the JT peak at 2.5 eV in Takenaka {\it et al.}'s results, 
we obtain $U$=1.6 eV.
The obtained value of $U$ is close to the difference 
of the experimental peak position and 
the calculated peak position for the $U$=0 case.
The room temperature results are shown in
Figs. 6(b) and 6(d). 
As we increase the value of $U$, 
the peak position shifts upwards by $\approx$ $U$, 
and the peak intensity decreases.
For $U$=1.6 eV, the calculated maximum peak intensity is 
730 $\Omega^{-1} {\rm cm}^{-1}$, width 1.2 eV,
and $K_{\rm av,JT}^{300K} \approx$ 0.145 eV.
Therefore, the calculated spectral weight is close to
the observed spectral weight.
With these determined values of $U$,
we calculate $T$=0 K results shown in Figs. 6(a) and 6(c),
from which we can see the anisotropy and temperature 
dependence of optical conductivity.

In the no-hopping case, the energy levels on each site are
$ -\lambda_{\Uparrow}' \sqrt{\bar{v}^2+\bar{w}^2} $, 
$\lambda_{\Uparrow}'  \sqrt{\bar{v}^2+\bar{w}^2}$, 
$2 J_{\rm H}' S_{\rm c} - \lambda_{\Downarrow}' 
\sqrt{\bar{v}^2+\bar{w}^2}$,
and 
$2 J_{\rm H}' S_{\rm c} + 
\lambda_{\Downarrow}' \sqrt{\bar{v}^2+\bar{w}^2}$.
Even though the finite hopping gives dispersion to the energy levels, 
the peak positions are close to the energy differences 
between different levels. 
Therefore, the JT peak position is 
close to $2\lambda_{\Uparrow}' \sqrt{\bar{v}^2+\bar{w}^2}$, 
the Hund peak position is close 
to $2 J_{\rm H}' S_{\rm c}+(\lambda_{\Uparrow}'-\lambda_{\Downarrow}')
\sqrt{\bar{v}^2+\bar{w}^2}$,
and the
JT+Hund peak position is close to 
$2 J_{\rm H}' S_{\rm c}+(\lambda_{\Uparrow}'
+\lambda_{\Downarrow}')\sqrt{\bar{v}^2+\bar{w}^2}$.
To see the effect of different type of Coulomb interaction 
and other parameters, we use different values
of $\lambda_{\Downarrow}'$, $J_{\rm H}'S_{\rm c}$ 
and calculate optical conductivities 
without changing the value of $\lambda_{\Uparrow}'$.
We observe that the JT peak position and spectral weight 
do not change very much.
Therefore even if we use a more general 
Coulomb interaction, with  $U_{(\alpha,\beta),(a,b)}$
dependent on the indices, 
as far as we fix the JT peak position by fixing $\lambda_{\Uparrow}'$, 
its spectral weight does not change very much.

For the above model, we also calculate the variation 
of the level occupancies 
as a function of $U$.
The results for 
the lowest lying orbital
$<\hat{n}_{\Uparrow,-}>$ are shown in Fig. 7 for $T$=0 K and $T$=300 K,
which shows that   
as $U$ increases,
the $e_g$ electrons are more likely to have
spins parallel to the core spins and 
stay in the ground state of the local lattice distortion.
The curves however show that the value of 
$U$ required to fit Takenaka {\it et al.}'s data
does not change the ground state occupancy much.

\section{Comparison with LSDA calculation of optical conductivity}
Terakura {\it et al.} have calculated the $T$=0 K 
optical conductivity
using
optical matrix elements and energies
obtained from their LSDA band calculation. \cite{Terakura99}
Their conductivity is strikingly different 
from ours in two respects.
First, the form is different:
the sharp peaks we find are absent in their calculation.
We suspect that the difference is due in large part to
the 
0.01 Ry $ \approx$ 0.14 eV
level broadening
employed in Ref. \onlinecite{Terakura99}.
Figure 8 shows the effect of introducing an artificial 
broadening, $\epsilon$, into our calculation;
the result is to be compared with Fig. 2(a).
This figure shows that as broadening is increased, 
the peaks diminish in amplitude and become more
symmetrical (although there is always more asymmetry in
our calculation than in Ref. \onlinecite{Terakura99}).
Particularly, the Hund peak in $\sigma_{zz}$ 
is expected to have additional  
broadening of about 0.17 eV as mentioned in Section III.
Therefore,
if we include 0.01 Ry broadening assumed in 
Ref. \onlinecite{Terakura99}, the total broadening for the Hund peak
in $\sigma_{zz}$ will be 
$\epsilon \approx$ 0.3 eV, which explains the absence of this peak
feature in Ref. \onlinecite{Terakura99}.
Therefore, we believe the level broadening along with the presence,
in the calculation of Ref.\ \onlinecite{Terakura99},
of other bands, accounts for the difference in line shape.

A far more serious discrepancy is the difference in spectral weights.
The area under the lowest conductivity peak 
in Ref.\ \onlinecite{Terakura99}
is about a factor of 4 smaller than in our calculation.
This difference seems not to be caused by a trivial error:
in our calculation,
the kinetic energies obtained from Eq.\ (\ref{eq:KE}), 
from Eq.\ (\ref{eq:sum}) (i.e., direct integration of $\sigma$), 
or from the Hellman-Feynman theorem 
agree.
According to Hellman-Feynman theorem, 
$K=K_{xx}+K_{yy}+K_{zz}$ can be found 
from the ground state energy $E_0/N_{Mn}$ by 
\begin{equation}
K=-t_0 \frac{ d (E_{0}/N_{Mn} ) }{d t_0 }, \label{eq:Hellman}
\end{equation}
where $t_0$ is the hopping parameter defined in Section II.
At $T$=0 K, we have calculated
$E_0$ as the sum of energies of the filled bands,
\begin{equation}
\frac{E_0}{N_{Mn}}=\frac{2}{(2 \pi )^3}
\int_{R} d \vec{k}
\left[ E_1 ( \vec{k} ) + E_2 ( \vec{k} ) \right]
\end{equation}
and evaluated $K$ using Eq. (\ref{eq:Hellman}).
The result obtained in this way is consistent with the results
in Table I.

We examine the size of the possible error due to the following
two approximations we have made: First, we have assumed that the hopping
between Mn ions, which originates from Mn-O hopping,
can be effectively represented without
explicit consideration of O band.
Second, we have used tight binding approximation.

To study the 
effects of Mn-O hybridization 
on the conductivity in the dominantly Mn bands,
we consider a simple model of a
1-d Mn-O chain along $x$ direction.
Each unit cell contains one Mn ion at position $R^{\rm Mn}_i=n_i a_0 $
with a d-orbital represented by $d^{\dagger}_i$,
and one oxygen ion at position $R^{\rm O}_i=(n_i+1/2) a_0$ with 
a p-orbital represented by $p^{\dagger}_i$.
We consider a Mn-O hopping of magnitude $t_{\rm Mn-O}$ and choose sign to
reflect the symmetry of the O p-orbital
(the sign can be removed by change of k space origin).
In addition, to model the Jahn-Teller distortion,
we consider alternating periodic potential $\Delta$ 
on the Mn site. We represent the energy of the d-level relative to the
p-level by $V$. For simplicity we assume spinless electrons.
This can be represented by the following Hamiltonian.
\begin{eqnarray}
H&=&-\frac{t_{\rm Mn-O}}{2} \sum_i
\left(
d_{i}^{\dagger}p_{i}
-p_{i}^{\dagger} d_{i+1}
-d_{i+1}^{\dagger} p_{i}
+p_{i}^{\dagger} d_{i} +H.c. \right) \nonumber \\
& &+ \sum_{i} \frac{\Delta}{2}(-1)^i d_{i}^{\dagger} d_{i}
- V p_{i}^{\dagger} p_{i} . 
\end{eqnarray}
We obtain the exact band structure and optical
conductivity and compare this to the band structure and conductivity
obtained from nearest-neighbor
tight-binding fit to the two uppermost (Mn-dominant) bands.
The difference turns out small. 
The effective tight binding Hamiltonian is 
\begin{equation}
H_{\rm eff}=-\frac{t_{\rm eff}}{2} \sum_i 
\left( d_{i}^{\dagger} d_{i+1} +
d_{i}^{\dagger} d_{i-1} +H.c. \right)
+\sum_i \frac{\Delta_{\rm eff}}{2} (-1)^i d_{i}^{\dagger} d_{i} 
\label{eq:TB}
\end{equation}
By transforming into $k$ space, we can find band structure for $H$ and
$H_{\rm eff}$.
The band structure of $H_{\rm eff}$ is simply given by 
\begin{equation}
E= \pm \sqrt{4 t_{\rm eff}^2 \cos^2 k +\frac{\Delta_{\rm eff}^2}{4}}
\end{equation}
For given $t_{\rm Mn-O}$, $\Delta$, and $V$, we can fit the band structure
of $H_{\rm eff}$ to that of $H$ to determine $t_{\rm eff}$ and
$\Delta_{\rm eff}$.

For $t_{\rm Mn-O}$=2.0 eV, $\Delta$=1.0 eV, 
we try $V$ = 1.0 eV and 10 eV.
The obtained d-bands are shown in Figs. 9(a) and 9(b)
($V$=1.0 eV and 10 eV, respectively) as
solid lines, 
along with the best tight
binding fit as dotted lines.
Fitted parameter values are $t_{\rm eff}$ =0.88 eV, 
$\Delta_{\rm eff}$ =0.59 eV for $V$=1.0 eV, and $t_{\rm eff}$=0.35 eV,
$\Delta_{\rm eff}$=0.93 eV for $V$=10 eV.
It shows that when $V$ =1.0 eV, the fitting has error of 10 \%
of d-band width 
(comparable to the error in the fits used in Section II - IV),
but when $V$=10 eV, the fitting has negligible error.
For these two cases, we assume half filling, and calculate optical
conductivities
shown in Figs. 9(c) and 9(d).
The insets show the integrated spectral weight,
$K(\hbar \omega) = ( 2 \hbar^2 a_0 / \pi e^2) \int_0^{\hbar \omega}
\sigma ( \omega ) d\omega   $.
In this calculation, we assumed the Mn-Mn distance is $a_0$=4.034
$\AA$, 
and the cross sectional area perpendicular to the direction of the
chain is $a_0^2$.
For $V$ =10 eV, the two calculations give almost identical results.
For $V$=1.0 eV, the tight binding fit has about 25 \% larger spectral
weight.
We therefore expect our Mn-only approximation yields errors
$\approx$ 25 \%.

Next, to estimate the error of the tight binding approximation
(i.e., of the Peierls approximation to the optical
matrix elements),
we consider the following Kronig-Penney model.
\begin{equation}
\hat{H}=-\frac{1}{2}\frac{d^2}{dx^2}
+\sum_{n=-\infty}^{\infty} \left
[
\frac{\lambda_1}{2}\delta(x-2n-1)+\frac{\lambda_2}{2}\delta(x-2n)\right]
\label{eq:KP}
\end{equation}
where $\lambda_1$, $\lambda_2 < $ 0, $ \hbar $=$m_e$=$e$=1, and spinless
electrons are assumed for simplicity.
Once the values of $\lambda_1$ and $\lambda_2$ are given, we can find the
band structure,
the eigenstates $\psi_k(x)$,
and the optical conductivity,
and compare this to the Peierls approximation.
The band structure has two bands;
at n=1/2, one is filled and the other is empty. 
Therefore,
we can calculate the optical conductivity
via
\begin{equation}
\sigma = \frac{1}{\omega}\frac{1}{2\pi}
\int_{-\frac{\pi}{2}}^{\frac{\pi}{2}}
\frac{ \left| <2,k|\frac{d}{dx}|1,k> \right|^2 \epsilon }
{ \left[ \omega -E_2(k) +E_1(k) \right]^2 +\epsilon^2 },
\end{equation}
where 1 and 2 are the band indices.
For $\lambda_1=-4 $, $\lambda_2=-5 $, we calculate the exact band
structure, shown in Fig. 10(a) along 
with the best tight binding fit
from Eq. (\ref{eq:TB})
($t_{\rm eff}$=1.33, $\Delta_{\rm eff}$=0.6).
It shows that the error is about 7 \% of the total band width.
Figure 10(b) shows the calculated optical conductivity for the exact K-P
model
and tight binding fit. 
The spectral weight of the tight binding fit is about 20 \% larger than
that of exact result, as shown in the inset.

In our calculation of optical conductivity, our band fitting has error
about 0.2 eV, which corresponds to about 5 \% of total 
Mn $e_g \uparrow $ band width.
So, we expect our approximation has similar size error in spectral
weight
as the two cases considered above.
Therefore, we expect that our calculated optical conductivity may have 
overestimated spectral weight by about 20 \%, but not 400 \%.
Therefore, we believe that within this error, 
our approximations are valid. 

The relation between the kinetic energy and the optical
spectral weight follows from the two assumptions of 
gauge invariance and reasonably localized d-electron
wave functions.
The success of the tight binding fit
confirms this localized character.
A tight binding parameterization of the band structure has been
used to study $\sigma(\omega)$ in other correlated electron
contexts,\cite{Dagotto94,Baeriswyl86}
and works well in high $T_c$ case.
The apparent discrepancy found here for the manganite,
is an important issue
for future research.

\section{Conclusion}
We have calculated the optical conductivity of ${\rm LaMnO_3}$
and have shown that the available data 
are consistent with the band theory estimation for hopping
parameter 
and $e_g$ level
splitting.
Our main prediction for the change in functional form and 
magnitude of $\sigma$ as $T$ is decreased
below $T_{\rm N}$
is contained in Fig. 6.
Experimental determination of the Hund's coupling 
as well as final validation of the model must await 
definitive
measurements of the magnitude and 
temperature dependence of $\sigma(\omega)$.
 
In conclusion, we comment on the implications of our results.
First, we note that the estimate 
of the electron-lattice coupling derived from 
band theory is 
in good agreement with that derived directly from crystallography data.
Second, we observe that the electron-lattice interaction by itself
does not account for the magnitude of the gap or the spectral weight 
in the absorption spectrum.
A Coulomb interaction $U$ $\approx$ 1.6 eV
is also required.
This value puts ${\rm LaMnO_3}$ in the weak-intermediate coupling range:
the Coulomb interaction is approximately 40\% 
of the full band width 6 $t_o \approx$ 3.6 eV.
In the simple one-band Hubbard model,
a Coulomb interaction of this size (relative to the band width)
does not significantly 
affect properties (such as optical spectral weights)
at reasonable dopings of order 0.2 or larger.
The effects of this moderate Coulomb coupling on properties of models
of doped manganites deserve further attention.
Many authors have argued on the basis of photoemission data 
\cite{Saitoh97} that
the Coulomb repulsion is large (5-10 eV); however 
as noted by the experimentalists themselves, because the
manganites are charge-transfer rather than Mott-Hubbard materials
(as are the high-Tc superconductors) the $U$ measured in photoemission
is not directly relevant to the low ($\hbar\omega <$ 4 eV) energy physics
of interest here.

Our data analysis focuses on robust features 
(peak positions and spectral weights)
and is insensitive to the approximations we made.
Uncertainties in the tight binding parameterization
of the band structure leads to
an error$\approx$ 0.2 eV 
in peak position, which
is not important here;
the consistency of the peak position and spectral weights
leads us to believe the band theory
estimates 
of $t_o$ are reasonably accurate.
Uncertainties in the estimates
of electron-phonon coupling $\lambda$ 
could change our estimated Coulomb repulsion by 
around 0.2 eV.
We note however that we have not 
included any excitonic effects arising from
the first neighbor interactions
such as those proposed by Maekawa {\it et al.}.\cite{Maekawa98}
At $T$=0 K, the two-dimensional 
character and general flatness of the bands suggest that 
these might be important and interesting to look for.

The crucial prediction of the present model 
is the dramatic change in optical absorption
with temperature.
This change is a robust feature of the model,
and comes from 
a dramatic shift in spectral weight caused by 
ferromagnetic spin ordering,
along with a very nearly two dimensional character of the bands
at $T \rightarrow$ 0,
caused by the between-planes antiferromagnetism.
Early data \cite{Okimoto97} reported 
only a weak temperature dependence of the optical absorption;
if these data are reproduced,
then
our fundamental picture of the manganites 
based on $e_g$ electron with 
electron-lattice and electron-electron interactions
must be modified. 

Finally, we note that a troubling discrepancy with band theory
calculations exists.
Further work is needed to find origin 
of the differences.

We thank H. Drew, S. Louie, O. Myrasov, 
and M. Quijada  for helpful discussions,
and NSF-DMR-9705182 and the University of Maryland MRSEC for support.

\appendix
\section{}
Without considering the Coulomb repulsion, 
the total Hamiltonian at $T$=0 K is
given by the following expression.
\begin{equation}
H_{\rm tot}=\sum_{\alpha=\uparrow,\downarrow, \vec{k} \in R}
d_{\alpha,\vec{k}}^{\dagger} H_{\alpha,\vec{k}} d_{\alpha,\vec{k}},
\end{equation}
where
\begin{eqnarray}
d_{\alpha,\vec{k}}^{\dagger}&=& 
(d_{1,\vec{k},\alpha}^{\dagger}, 
d_{2,\vec{k},\alpha}^{\dagger},
d_{1,\vec{k}+(\pi,\pi,0),\alpha}^{\dagger},
d_{2,\vec{k}+(\pi,\pi,0),\alpha}^{\dagger}, \nonumber \\
& &d_{1,\vec{k}+(0,0,\pi),\alpha}^{\dagger},
d_{2,\vec{k}+(0,0,\pi),\alpha}^{\dagger},
d_{1,\vec{k}+(\pi,\pi,\pi),\alpha}^{\dagger},
d_{2,\vec{k}+(\pi,\pi,\pi),\alpha}^{\dagger}),  \\
H_{\vec{k},\alpha}&=&\left(
\begin{array}{cccc}
M_1+G+V  &  W  &  G_{\alpha}  &  0  \\
W & M_2+G+V & 0 &  G_{\alpha}   \\
G_{\alpha} & 0 & M_3+G+V & W \\
0 & G_{\alpha} & W & M_4+G+V 
\end{array} \right), \\
M_1&=&\left( \begin{array}{cc}
  -\frac{t_0}{2}(\cos k_x + \cos k_y+4 \cos k_z ) 
&  \frac{\sqrt{3}t_0}{2}(\cos k_x -\cos k_y)    \\
   \frac{\sqrt{3}t_0}{2}(\cos k_x -\cos k_y)  & 
    -\frac{3 t_0}{2}(\cos k_x + \cos k_y)    
\end{array} \right), \\
M_2&=&\left( \begin{array}{cc}
    \frac{t_0}{2}(\cos k_x + \cos k_y-4 \cos k_z ) & 
 -\frac{\sqrt{3}t_0}{2}(\cos k_x -\cos k_y)    \\
    -\frac{\sqrt{3}t_0}{2}(\cos k_x -\cos k_y) 
 &  \frac{3 t_0}{2}(\cos k_x + \cos k_y)    
\end{array} \right), \\
M_3&=&\left( \begin{array}{cc}
    -\frac{t_0}{2}(\cos k_x + \cos k_y-4 \cos k_z ) &
  \frac{\sqrt{3}t_0}{2}(\cos k_x -\cos k_y)    \\
  \frac{\sqrt{3}t_0}{2}(\cos k_x -\cos k_y) 
 & -\frac{3 t_0}{2}(\cos k_x + \cos k_y)    
\end{array} \right), \\
M_4&=&\left( \begin{array}{cc}
   \frac{t_0}{2}(\cos k_x + \cos k_y+4 \cos k_z ) &
  -\frac{\sqrt{3}t_0}{2}(\cos k_x -\cos k_y)    \\
  -\frac{\sqrt{3}t_0}{2}(\cos k_x -\cos k_y)  &
  \frac{3 t_0}{2}(\cos k_x + \cos k_y)    
\end{array} \right), \\
G&=&\left( \begin{array}{cc}
                   J_{\rm H}S_{\rm c} &  0    \\
                       0  &  J_{\rm H}S_{\rm c}   
\end{array} \right), \\
G_{\uparrow}&=&-G,      \\
G_{\downarrow}&=&G,     \\
V&=&\left( \begin{array}{cc}
                   \lambda\bar{v} &  0    \\
                       0  & -\lambda\bar{v}    
\end{array} \right),  \\
W&=&\left( \begin{array}{cc}
                 0   &   -\lambda\bar{w}     \\
                  -\lambda\bar{w}  &  0    
\end{array} \right),  \\
R&=&\{\vec{k} | |k_x|+|k_y| < \pi \ {\rm and}\  |k_z|  < \pi/2 \}.  
\end{eqnarray}

\section{}
Without considering the Coulomb repulsion, 
the effective total Hamiltonian at $T$=300 K is
given by the following expression.
 \begin{equation}
H^{\rm eff, 300 K}_{\rm tot}=\sum_{\vec{k} \in S}
d_{\vec{k}}^{\dagger} H^{\rm eff, 300K}_{\vec{k}} d_{\vec{k}},
\end{equation}
where
\begin{eqnarray}
d_{\vec{k}}^{\dagger}&=& 
(d_{1,\vec{k},\Uparrow}^{\dagger}, 
d_{2,\vec{k},\Uparrow}^{\dagger},
d_{1,\vec{k}+(\pi,\pi,0),\Uparrow}^{\dagger},
d_{2,\vec{k}+(\pi,\pi,0),\Uparrow}^{\dagger}, \nonumber \\
& &
d_{1,\vec{k},\Downarrow}^{\dagger}, 
d_{2,\vec{k},\Downarrow}^{\dagger},
d_{1,\vec{k}+(\pi,\pi,0),\Downarrow}^{\dagger},
d_{2,\vec{k}+(\pi,\pi,0),\Downarrow}^{\dagger}
),  \\
H^{\rm eff, 300K}_{\vec{k}}&=&\left(\begin{array}{cccc}
\frac{2}{3}M_1+V & W & \frac{2}{3}M_1 & 0 \\
W & \frac{2}{3}M_2+V & 0 & \frac{2}{3}M_2 \\ 
\frac{2}{3}M_1 & 0 & \frac{2}{3}M_1+V+2 G & W \\
0 & \frac{2}{3}M_2 & W & \frac{2}{3}M_2+V+2 G 
\end{array} \right), \\
S&=&\{\vec{k} | |k_x|+|k_y| < \pi \  {\rm and} \ |k_z|  < \pi \}.  
\end{eqnarray}
$M_1$, $M_2$, $V$, $W$ and $G$ are defined in Appendix A.

\section{}
At $T$=0 K, the total mean field Hamiltonian with the Coulomb interaction,
\begin{equation}
H^{\rm MF}_{\rm tot}= 
\sum_{\alpha=\uparrow,\downarrow, \vec{k} \in R}
d_{\alpha,\vec{k}}^{\dagger} 
H_{\alpha,\vec{k}}^{\rm MF} d_{\alpha,\vec{k}},
\end{equation}
has the same form as $H_{\rm tot}$ in the Appendix A 
with $J_{\rm H} S_{\rm c} \rightarrow J_{\rm H}' S_{\rm c}$, 
$\lambda \rightarrow \lambda_{av}'$
and the following additional term.
\begin{equation}
H_{\rm add,T=0K}^{\rm MF}=
\sum_{\alpha=\uparrow,\downarrow, \vec{k} \in R}
d_{\alpha,\vec{k}}^{\dagger} 
H_{{\rm add},\alpha,\vec{k}}^{\rm MF} d_{\alpha,\vec{k}},
\end{equation}
where
\begin{equation}
H_{{\rm add},\alpha,\vec{k}}^{\rm MF}=\left(\begin{array}{cccc}
0  &  0  &  \delta \bar{V}_{\alpha}  &  \delta \bar{W}_{\alpha}   \\
0  &  0  &  \delta \bar{W}_{\alpha} &  \delta \bar{V}_{\alpha}   \\
 \delta \bar{V}_{\alpha} &  \delta \bar{W}_{\alpha} & 0 & 0 \\
 \delta \bar{W}_{\alpha} &  \delta \bar{V}_{\alpha} & 0 & 0 
\end{array} \right), \\
\end{equation}
\begin{eqnarray}
\delta \bar{V}_{\alpha}&=&\left( \begin{array}{cc}
                   \delta \lambda_{\alpha}' \bar{v} &  0    \\
                       0  & -\delta \lambda_{\alpha}' \bar{v}    
\end{array} \right),  \\
\delta \bar{W}_{\alpha}&=&\left( \begin{array}{cc}
                       0  &  - \delta \lambda_{\alpha}' \bar{w}     \\
                       - \delta \lambda_{\alpha}' \bar{w}  &  0    
\end{array} \right),   \\
\delta \lambda_{\uparrow}' &=& \delta \lambda' , \\
\delta \lambda_{\downarrow}' &=& -\delta \lambda' .  
\end{eqnarray}

Similarly, for $T$=300 K 
the effective mean field total Hamiltonian 
with the Coulomb interaction,
\begin{equation} 
H^{\rm MF, 300K}_{\rm tot}=
\sum_{\vec{k} \in S}
d_{\vec{k}}^{\dagger} 
H_{\vec{k}}^{\rm MF, 300K} d_{\vec{k}},
\end{equation}
has the same form as $H_{\rm tot}^{\rm eff, 300K}$ in the Appendix B 
with  $J_{\rm H} S_{\rm c} \rightarrow J_{\rm H}' S_{\rm c}$, 
$\lambda \rightarrow \lambda_{av}'$,
and the following additional term.
\begin{equation}
H_{\rm add}^{\rm MF, 300K}=\sum_{\vec{k} \in S}
d_{\vec{k}}^{\dagger} 
H_{{\rm add},\vec{k}}^{\rm MF, 300K} d_{\vec{k}},
\end{equation}
where
\begin{equation}
H_{{\rm add},\vec{k}}^{\rm MF, 300K}=\left(\begin{array}{cccc}
\delta \bar{V}  &  \delta \bar{W} & 0 & 0   \\
\delta \bar{W} &  \delta \bar{V} & 0 & 0   \\
 0 & 0  & -\delta \bar{V} &  -\delta \bar{W}  \\
 0 & 0  &  -\delta \bar{W} & - \delta \bar{V}  
\end{array} \right). 
\end{equation}
\begin{eqnarray}
\delta \bar{V}&=&\left( \begin{array}{cc}
                   \delta \lambda' \bar{v} &  0    \\
                       0  & -\delta \lambda' \bar{v}    
\end{array} \right),  \\
\delta \bar{W}&=&\left( \begin{array}{cc}
                       0  &  - \delta \lambda' \bar{w}     \\
                       - \delta \lambda' \bar{w}  &  0    
\end{array} \right).   
\end{eqnarray}

With these Hamiltonians for given $U_{(\alpha,\beta),(a,b)}$,
we can repeatedly calculate 
$<\hat{n}_{\beta,b}>$ until its value converges.
At $T$=0 K, $<\hat{n}_{\alpha,a}>$
is 
given by the following  expression.
\begin{equation}
<\hat{n}_{\alpha,a}> = 
\frac{1}{(2 \pi)^3} \int_R d \vec{k}
\sum_{j=1,2} 
[P_{\uparrow}(\vec{k})^{\dagger}  
Q_{\uparrow \alpha a} P_{ \uparrow}(\vec{k})]_{jj} +
[P_{\downarrow}(\vec{k})^{\dagger}  
Q_{\downarrow \alpha a} P_{\downarrow} (\vec{k})]_{jj},  
\end{equation}
where
\begin{eqnarray}
Q_{\uparrow \Uparrow,\pm}&=&Q_{\downarrow \Downarrow,\pm}=Q_{1\pm}, \\
Q_{\downarrow \Uparrow,\pm}&=&Q_{\uparrow \Downarrow,\pm}=Q_{2\pm}, \\
Q_{1\pm}&=&
\left(\begin{array}{cccc}
A_{\pm}  & B_{\pm}  &  A_{\pm}  &  B_{\pm}  \\
B_{\pm} & A_{\pm} & B_{\pm} &  A_{\pm}   \\
A_{\pm} & B_{\pm} & A_{\pm} & B_{\pm} \\
B_{\pm} & A_{\pm} & B_{\pm} & A_{\pm} 
\end{array} \right), \\
Q_{2\pm}&=&
\left(\begin{array}{cccc}
A_{\pm}  & B_{\pm}  &  -A_{\pm}  &  -B_{\pm}  \\
B_{\pm} & A_{\pm} & -B_{\pm} &  -A_{\pm}   \\
-A_{\pm} & -B_{\pm} & A_{\pm} & B_{\pm} \\
-B_{\pm} & -A_{\pm} & B_{\pm} & A_{\pm} 
\end{array} \right), \\
A_{\pm}&=&\left(\begin{array}{cc}
\alpha_{\pm}^2/2   &  0 \\
0  &  \beta_{\pm}^2/2  
\end{array} \right), \\
B_{\pm}&=&\left(\begin{array}{cc}
0 & \alpha_{\pm} \beta_{\pm}/2    \\
\alpha_{\pm} \beta_{\pm}/2    &  0
\end{array} \right), \\
\alpha_{\pm}&=&\frac{\bar{w}}{\sqrt{2(\bar{v}^2+\bar{w}^2) 
\mp 2 \bar{v} \sqrt{\bar{v}^2+\bar{w}^2}} }, \\
\beta_{\pm}&=&\frac{\bar{v}\mp\sqrt{\bar{v}^2+\bar{w}^2}}
{\sqrt{2(\bar{v}^2+\bar{w}^2) 
\mp  2 \bar{v} \sqrt{\bar{v}^2+\bar{w}^2} } },
\end{eqnarray}
and $P_{\alpha}(\vec{k})$ is the matrix diagonalizing 
$H_{\alpha,\vec{k}}^{\rm MF}$.

At $T$=300 K, 
\begin{equation}
<\hat{n}_{\alpha,a}> = 
\frac{1}{(2 \pi)^3}
\int_S  d \vec{k}
\sum_{j=1,2} 
[P'(\vec{k})^{\dagger}  Q'_{ \alpha a} P'(\vec{k})]_{jj},  
\end{equation}
where
\begin{eqnarray}
Q'_{\Uparrow \pm}&=&
\left(\begin{array}{cccc}
A_{\pm}  & B_{\pm}  &  0  &  0  \\
B_{\pm} & A_{\pm} & 0 &  0   \\
0 & 0 & 0 & 0 \\
0 & 0 & 0 & 0 
\end{array} \right), \\
Q'_{\Downarrow \pm}&=&
\left(\begin{array}{cccc}
 0 & 0  &  0  &  0  \\
 0 & 0 & 0 &  0   \\
0 & 0 & A_{\pm}  & B_{\pm} \\
0 & 0 & B_{\pm} & A_{\pm} 
\end{array} \right),
\end{eqnarray}
and  $P'(\vec{k})$ is the matrix diagonalizing 
$H_{\vec{k}}^{\rm MF, 300K}$.

\newpage

\begin{figure}
\caption{
Fitted $e_g$ band structure of ${\rm LaMnO_3}$ :
$t_0$ = 0.622 eV, $2 J_{\rm H}S_{\rm c}$ = 2.47 eV, 
$\lambda$=1.38 eV/$\AA$, and $\mu$=0.4 eV. 
($\pi$,0,0), (0,0,0), ($\pi$/2,$\pi$/2,$\pi$/2), 
and ($\pi$,0,$\pi$/2) points correspond to
M, $\Gamma$, R, and A points in Ref.
11 
respectively. 
}
\label{fig1}
\end{figure}

\begin{figure}
\caption{
Optical conductivities $\sigma_{xx}$ (solid lines) 
and $\sigma_{zz}$(dotted lines)
for $t_0$ = 0.622 eV, and $2 J_{\rm H}S_{\rm c}$ = 2.47 eV
without Coulomb repulsion(U=0).
}   
\label{fig2}
\end{figure}

\begin{figure}
\caption{
Optical conductivities $\sigma_{xx}$(solid lines) 
and $\sigma_{zz}$(dotted lines)
for $t_0$ = 0.622 eV, and $2 J_{\rm H}S_{\rm c}$ = 2.10 eV
without Coulomb repulsion(U=0).
}   
\label{fig3}
\end{figure}

\begin{figure}
\caption{
Optical conductivities $\sigma_{xx}$ (solid lines) 
and $\sigma_{zz}$(dotted lines)
for $t_0$ = 0.622 eV, and $2 J_{\rm H}S_{\rm c}$ = 2.47 eV,
$\lambda$ = 1.38 $eV/\AA$, U=0 with different orbital ordering angle 
$\theta $.
}
\label{fig4}
\end{figure}

\begin{figure}
\caption{
Average optical conductivities $\sigma_{av}$
for $t_0$ = 0.622 eV and $2 J_{\rm H}S_{\rm c}$ = 2.47 eV at T=300 K
without Coulomb repulsion(U=0).
}   
\label{fig5}
\end{figure}

\begin{figure}
\caption{
Optical conductivities 
for $t_0$ = 0.622 eV, $2 J_{\rm H}S_{\rm c}$ = 2.47 eV, 
$\lambda$=1.38 eV/$\AA$,
and  
U=1.59 eV.
}   
\label{fig6}
\end{figure}

\begin{figure}
\caption{
Occupancies of the lowest-lying orbital versus U for 
$t_0$ = 0.622 eV, $2 J_{\rm H}S_{\rm c}$ = 2.47 eV, 
and $\lambda$=1.38 eV/$\AA$
at $T$=0 K and 300 K.
}   
\label{fig7}
\end{figure}

\begin{figure}
\caption{
Optical conductivities $\sigma_{xx}$ (solid lines) 
and $\sigma_{zz}$(dotted lines)
for $t_0$ = 0.622 eV, and $2 J_{\rm H}S_{\rm c}$ = 2.47 eV,
$\lambda$ = 1.38 $eV/\AA$, U=0 with different broadening $\epsilon $.
}
\label{fig8}
\end{figure}

\begin{figure}
\caption{
Mn d band structures[(a),(b)] and 
optical conductivities[(c),(d)] 
of Mn-O chain in a model explicitly 
considering O p-level (solid lines) and in the best-fit effective
Mn-Mn chain model (dotted lines) for $t_{Mn-O}$=2.0 eV,
$\Delta $=1.0 eV, and
$V$ = 1.0 eV, $t_{\rm eff}$=0.88 eV , $\Delta_{\rm eff}$= 0.59 eV
[(a),(c)], 
and  $V$=10 eV, $t_{\rm eff}$= 0.35 eV, $\Delta_{\rm eff}$= 0.93 eV
[(b),(d)].
The insets show the integrated spectral weight represented in terms of
kinetic energy versus photon energy.
}
\label{fig9}
\end{figure}

\begin{figure}
\caption{
Band structures[(a)] and optical conductivities[(b)] 
for exact K-P model[Eq. (\ref{eq:KP})] with $\lambda_1=-4$,
$\lambda_2=-5 $
(solid line)
and for the  best tight binding fit(dotted line), 
$t_{\rm eff}$=1.33,
$\Delta_{\rm eff}$=0.6.
The inset shows the integrated spectral weight represented in terms of
kinetic energy versus photon energy.
}
\label{figure10}
\end{figure}

\newpage
\begin{table}
\caption{
Total spectral weight 
for $t_0$=0.622 eV without Coulomb interaction($U$=0),
expressed in terms of $K_x$ and $K_z$.
}

\label{table1}
\begin{tabular}{c|c||c|c|c|c}
$2 J_{\rm H}S_{\rm c}$(eV)  & $\lambda$(eV/$\AA$) &  
$K_{x}^{0 K}$(eV) & $K_{z}^{0 K}$(eV) 
&  $K_{x}^{300 K}$(eV) & $K_{z}^{300 K}$
(eV)  \\  \hline
2.10 & 1.38 &  0.295 & 0.242 & 0.290 & 0.211 \\
 2.10 & 2.20 &  0.236 & 0.206 & 0.248 & 0.179 \\
2.10 & 2.90 &  0.196 & 0.184 & 0.219 & 0.155 \\
\hline
2.47 & 1.38 &  0.294 & 0.217 & 0.277 & 0.199 \\
 2.47 & 2.20 &  0.235 & 0.185 & 0.236 & 0.167 \\
2.47 & 2.90 &  0.195 & 0.165 & 0.207 & 0.144 \\ 
\end{tabular}
\end{table}
\end{document}